\documentclass[aps,showpacs,superscriptaddress,preprint]{revtex4}%
\usepackage{amsfonts}
\usepackage{amsmath}
\usepackage{amssymb}
\usepackage{graphicx}%
\providecommand{\U}[1]{\protect\rule{.1in}{.1in}}

\begin{document}
\title{Equation of states and transport properties of warm dense beryllium:
A quantum molecular dynamics study}
\author{Cong Wang}
\affiliation{Institute of Applied Physics and Computational
Mathematics, P.O. Box 8009, Beijing 100088, People's Republic of
China}
\affiliation{Center for Applied Physics and Technology,
Peking University, Beijing 100871, People's Republic of China}
\author{Yao Long}
\affiliation{Institute of Applied Physics and Computational
Mathematics, P.O. Box 8009, Beijing 100088, People's Republic of
China}
\author{Ming-Feng Tian}
\affiliation{Institute of Applied Physics and Computational
Mathematics, P.O. Box 8009, Beijing 100088, People's Republic of
China}
\author{Xian-Tu He}
\affiliation{Institute of Applied Physics and Computational
Mathematics, P.O. Box 8009, Beijing 100088, People's Republic of
China}
\affiliation{Center for Applied Physics and Technology,
Peking University, Beijing 100871, People's Republic of China}
\author{Ping Zhang}
\thanks{Corresponding author: zhang\_ping@iapcm.ac.cn}
\affiliation{Institute of Applied Physics and Computational
Mathematics, P.O. Box 8009, Beijing 100088, People's Republic of
China}
\affiliation{Center for Applied Physics and Technology,
Peking University, Beijing 100871, People's Republic of China}


\begin{abstract}
We have calculated the equation of states, the viscosity and
self-diffusion coefficients, and electronic transport coefficients
of beryllium in the warm dense regime for densities from 4.0 to 6.0
g/cm$^{3}$ and temperatures from 1.0 to 10.0 eV by using quantum
molecular dynamics simulations. The principal Hugoniot is accordant
with underground nuclear explosive and high power laser experimental
results up to $\sim$ 20 Mbar. The calculated viscosity and
self-diffusion coefficients are compared with the one-component
plasma model, using effective charges given by the average-atom
model. The Stokes-Einstein relationship, which presents the
relationship between the viscosity and self-diffusion coefficients,
is found to hold fairly well in the strong coupling regime. The
Lorenz number, which is the ratio between thermal and electrical
conductivities, is computed via Kubo-Greenwood formula and compared
to the well-known Wiedemann-Franz law in the warm dense region.

\end{abstract}
\maketitle

\section{INTRODUCTION}
\label{sec-introduction}

The nature of compressed matter is of considerable interest for many
fields of modern physics, including astrophysics \cite{Saumon1995},
inertial confinement fusion (ICF) \cite{Atzeni2004,Lindl1998}, and
other related fields \cite{Fortov2007}. Materials under a pressure
greater than a few Mbar can be drived into a strongly coupled,
partially ionized fluid states, which is defined as the so-called
warm dense matter (WDM). Theoretical modeling and experimental
detection of the high pressure behavior of WDM are of great
challenge and are being in extensive investigations. Among various
kinds of WDMs, warm dense beryllium (Be) is of particular current
interest. The equation of states (EOS) and transport properties of
Be are very important in ICF, due to its appearance in the ablator
of Deuterium-Tritium (D-T) capsule. The compressibility of the
capsule, laser absorption, and instability growth at the
fuel-ablator interface sensitively depend on the thermo-physical
properties of Be \cite{Lindl2004}.

The EOS of Be at shock Hugoniot up to $\sim$ 18 Mbar have been
accessed by strong shock waves generated by underground nuclear
explosives \cite{Ragan1982,Nellis1997}. Then, it is also possible to
probe similar pressure range in the laboratory by high intensity
laser \cite{Cauble1998}. Despite the success of these techniques in
detecting wide range EOS of typical matters, one should note that
after several decades, only seven Hugoniot points are available for
Be, and more experimental data are desired for building successful
theoretical models, such as interatomic potentials or chemical
models currently used in SESAME EOS \cite{Lyon1992}. As another
parameter in determining the EOS, temperature, which is difficult to
be measured in  nuclear explosion or high power laser experiments,
is still needed to be clarified. Apart from the EOS, the atomic
diffusion coefficients and fluid viscosity are key ingredients to
control hydrodynamic instabilities near interfaces
\cite{Robey2003,Betti1998,Hammel2008}. The electronic dynamic
conductivity, from which dielectric function can be obtained,
determines a series of interactions between laser and matters
\cite{Seka2008,Boehly2006}. Due to these key issues to be addressed,
therefore, the thermo-physical properties of Be in the warm dense
region are highly recommended to be understood in a systematic and
self-consistent way.

In the present work, quantum molecular dynamics (QMD) simulations
\cite{Bezkrovniy2004,Lenosky2000}, where electrons are fully quantum
mechanically treated by finite-temperature density functional theory
(FT-DFT), have been introduced to study warm dense Be. The EOS are
extracted from a series of NVT assemble sampling over different
densities and temperatures, then the Hugoniot curve is calculated
from the Rankine-Hugoniot relation. The self-diffusion coefficient
and viscosity have been computed from the trajectory by the velocity
and the stress tensor autocorrelation function. The dynamic
conductivity, from which the DC conductivity and electronic thermal
conductance are derived, has been obtained from Kubo-Greenwood
formula. The rest of the paper is organized as follows: In Sec.
\ref{sec-method}, we briefly describe the QMD simulations and
computational method in determining the atomic transport properties
and electronic dynamic conductivity; In Sec. \ref{sec-analysis},
discussions are presented for the EOS and transport properties; And
In Sec. \ref{sec-conclusion}, we get our conclusions.

\section{COMPUTATIONAL METHOD}
\label{sec-method}

\subsection{Quantum Molecular Dynamics}
Our QMD simulations employed the Vienna ab initio Simulation Package
(VASP) \cite{Kresse1993,Kresse1996}. A series of volume fixed
supercells including $N$ atoms, which are repeated periodically
throughout the space, form the elements of our calculations. After
Born-Oppenheimer approximation, electrons are quantum mechanically
treated through plane-wave FT-DFT. The interaction between electron
and ion is presented by a projector augmented wave (PAW)
pseudopotential. The exchange-correlation functional is determined
within Perdew-Burke-Ernzerhof generalized gradient approximation.
The ions move classically according to the forces from the electron
density and the ion-ion repulsion. The system is kept in local
thermodynamic equilibrium with the electron and ion temperatures
equal ($T_{e}=T_{i}$). The electronic temperature is kept through
Fermi-dirac distribution of the electronic states, and the ion
temperature is secured through Nos\'{e}-Hoover thermostat
\cite{Hunenberger2005}.

We have chosen 216 atoms in the unit cell with periodic boundary
condition. A range of densities from $\rho$ = 1.84 g/cm$^{3}$ to 6.0
g/cm$^{3}$ and temperatures from $T$ = 300 K to $T$ = 120000 K are
selected to highlight the principle Hugoniot regions. The
convergence of the thermodynamic quantities plays an important role
in the accuracy of QMD simulations. In the present work, a
plane-wave cutoff energy of 800 eV is employed in all simulations so
that the pressure is converged within 2\%. We have also checked the
convergence with respect to a systematic enlargement of the
\textbf{k}-point set in the representation of the Brillouin zone.
The correction of higher-order \textbf{k} points on the EOS data is
slight and negligible. In the molecular dynamics simulations, only
$\Gamma$ point of the Brillouin zone is included, while 4 $\times$ 4
$\times$ 4 Monkhorst-Pack scheme grid points are used in the
electronic structure calculations. The dynamic simulations have
lasted 10 $\sim$ 20 ps with time steps of 0.5 $\sim$ 1.0 fs
according to different conditions. For each pressure and
temperature, the system is equilibrated within 1 $\sim$ 2 ps. The
EOS data are obtained by averaging over the final 5 ps molecular
dynamic simulations.

\subsection{Transport properties}
The self-diffusion coefficient $D$ can either be calculated from the
trajectory by the mean-square displacement
\begin{equation}\label{D_R}
    D=\frac{1}{6t}\langle|R_{i}(t)-R_{i}(0)|^{2}\rangle,
\end{equation}
or by the velocity autocorrelation function
\begin{equation}\label{D_v}
    D=\frac{1}{3}\int_{0}^{\infty}\langle V_{i}(t)\cdot V_{i}(0)\rangle dt,
\end{equation}
where $R_{i}$ is the position and $V_{i}$ is the velocity of the
$i$th nucleus. Only in the long-time limit, these two formulas of
$D$ are formally equivalent. Sufficient lengths of the trajectories
have been generated to secure contributions from the velocity
autocorrelation function to the integral is zero, and the mean
mean-square displacement away from the origin consistently fits to a
straight line. The diffusion coefficient obtained from these two
approaches lie within 1 \% accuracy of each other, here, we report
the results from velocity autocorrelation function.

The viscosity
\begin{equation}\label{eta_lim}
    \eta=\lim_{t\rightarrow\infty}\bar{\eta}(t),
\end{equation}
has been computed from the autocorrelation function of the
off-diagonal component of the stress tensor \cite{Allen1987}
\begin{equation}\label{eta}
    \bar{\eta}(t)=\frac{V}{k_{B}T}\int_{0}^{t}\langle P_{12}(0)P_{12}(t')\rangle
    dt'.
\end{equation}
The results are averaged from the five independent off-diagonal
components of the stress tensor $P_{xy}$, $P_{yz}$, $P_{zx}$,
$(P_{xx}-P_{yy})/2$, and $(P_{yy}-P_{zz})/2$.

Different from the self-diffusion coefficient, which involves
single-particle correlations and attains significant statistical
improvement from averaging over the particles, the viscosity depends
on the entire system and thus needs very long trajectories so as to
gain statistical accuracy. To shorten the length of the trajectory,
we use empirical fits \cite{Kress2011} to the integrals of the
autocorrelation functions. Thus, extrapolation of the fits to
$t\rightarrow\infty$ can more effectively determine the basic
dynamical properties. Both of the $D$ and $\bar{\eta}$ have been fit
to the functional in the form of $A[1-\exp(-t/\tau)]$, where $A$ and
$\tau$ are free parameters. Reasonable approximation to the
viscosity can be produced from the finite time fitting procedure,
which also serves to damp the long-time fluctuations.

The fractional statistical error in calculating a correlation
function $C$ for molecular-dynamics trajectories \cite{Zwanzig1969}
can be given by
\begin{equation}\label{error}
    \frac{\vartriangle C}{C}=\sqrt{\frac{2\tau}{T_{traj}}},
\end{equation}
where $\tau$ is the correlation time of the function, and $T_{traj}$
is the length of the trajectory. In the present work, we generally
fitted over a time interval of [0, $4\tau-5\tau$].

\subsection{Dynamic Conductivity}

The key to evaluate the electrical transport properties is the
kinetic coefficients. They are calculated using the following
Kubo-Greenwood formulation:
\begin{equation} \label{kubo}
\hat{\sigma}(\epsilon)=\frac{1}{\Omega}\sum_{k,k'}|\langle\psi_{k}|\hat{v}|\psi_{k'}\rangle|^{2}\delta(\epsilon_{k}-\epsilon_{k'}-\epsilon),
\end{equation}
where $\langle\psi_{k}|\hat{v}|\psi_{k'}\rangle$ are the velocity
matrix elements, $\Omega$ is the volume of the supercell, and
$\epsilon_{k}$ are the electronic eigenvalues. The kinetic
coefficients $\mathcal{L}_{ij}$ in the Chester-Thellung version
\cite{Chester1961} are given by
\begin{equation} \label{coefficient}
\mathcal{L}_{ij}=(-1)^{i+j}\int
d\epsilon\hat{\sigma}(\epsilon)(\epsilon-\mu)^{(i+j-2)}(-\frac{\partial
f(\epsilon)}{\partial \epsilon}),
\end{equation}
where $\mu$ is the chemical potential and $f(\epsilon)$ is the
Fermi-Dirac distribution function. The electrical conductivity
$\sigma$ is obtained as
\begin{equation} \label{sigma}
\sigma=\mathcal{L}_{11},
\end{equation}
and electronic thermal conductivity $K$ is
\begin{equation} \label{thermal}
K=\frac{1}{T}(\mathcal{L}_{22}-\frac{\mathcal{L}_{12}^{2}}{\mathcal{L}_{11}}),
\end{equation}
where $T$ is the temperature. Equations (\ref{sigma}) and
(\ref{thermal}) are energy-dependent, then the electrical
conductivity and electronic thermal conductivity are obtained
through extrapolating to zero energy. In order to get converged
transport coe¡Àcients, ten independent snapshots, which are selected
during one molecular dynamic simulation at given conditions, are
selected to calculate electrical conductivity and electronic thermal
conductivity as running averages.

\section{RESULTS AND DISCUSSION}
\label{sec-analysis}

\subsection{The equation of states}

\begin{table}[tbh]
\caption{Expansion coefficients $A_{ij}$ for the internal energy $E$ (eV/atom).}%
\centering
\begin{tabular*}{12cm}{@{\extracolsep{\fill}}llllr}
\hline\hline $A_{i,j}$ & $j=0$ & $j=1$ & $j=2$ \\\hline
$i=0$     &   0.4369   &    0.5386  &   0.3149 \\
$i=1$     &  -0.7316   &    1.1936  &  -0.1149 \\
$i=2$     &   0.3215   &   -0.1241  &   0.0117 \\
\hline
\end{tabular*}
\label{table_energy}%
\end{table}

\begin{table}[tbh]
\caption{Expansion coefficients $B_{ij}$ for pressure $P$ (GPa).}%
\centering
\begin{tabular*}{12cm}{@{\extracolsep{\fill}}llllr}
\hline\hline $B_{i,j}$ & $j=0$ & $j=1$ & $j=2$ \\\hline
$i=0$     & -17.9930   &   10.2237  &  -0.7659 \\
$i=1$     & -62.4972   &   24.4101  &   0.3885 \\
$i=2$     &  38.7063   &   -0.2107  &  -0.0218 \\
\hline
\end{tabular*}
\label{table_pressure}%
\end{table}

Wide range EOS ($\rho$ from 4.0 to 6.0 g/cm$^{3}$ and temperatures
of 1 $\sim$ 10 eV) have been calculated according to QMD
simulations, and the internal energy $E$ (eV/atom) and pressure $P$
(GPa) are fitted by expansions in terms of density (g/cm$^{3}$) and
temperature (eV) as follows:
\begin{equation}\label{E_fit}
    E=\sum A_{ij}\rho^{i}T^{j},
\end{equation}
\begin{equation}\label{E_fit}
    P=\sum B_{ij}\rho^{i}T^{j}.
\end{equation}
The fitted coefficient for $A_{i,j}$and $B_{i,j}$ are summarized in
Tables \ref{table_energy} and \ref{table_pressure}. Here, the
internal energy $E$ at 1.84 g/cm$^{3}$ and $T$ = 300 K has been
taken as zero.

Based on the fitted EOS, the principal Hugoniot curve can be derived
from the Rankine-Hugoniot equation, which is the locus of points in
($E,P,V$)-space satisfying the condition
\begin{equation}\label{eq_RH}
    (E_{0}-E_{1})+\frac{1}{2}(V_{0}-V_{1})(P_{0}+P_{1})=0,
\end{equation}
where the subscripts 0 and 1 denote the initial and shocked state,
respectively. This relation follows from conservation of mass,
momentum, and energy for an isolated system compressed by a pusher
at a constant velocity. In the canonical (NVT) ensemble in which
both $E$ and $P$ are temperature dependent, the locus of states
which satisfies Eq. (\ref{eq_RH}) is the so-called principal
Hugoniot, which describes the shock adiabat between the initial and
final states.

\begin{figure}[!ht]
\includegraphics[width=12.0cm]{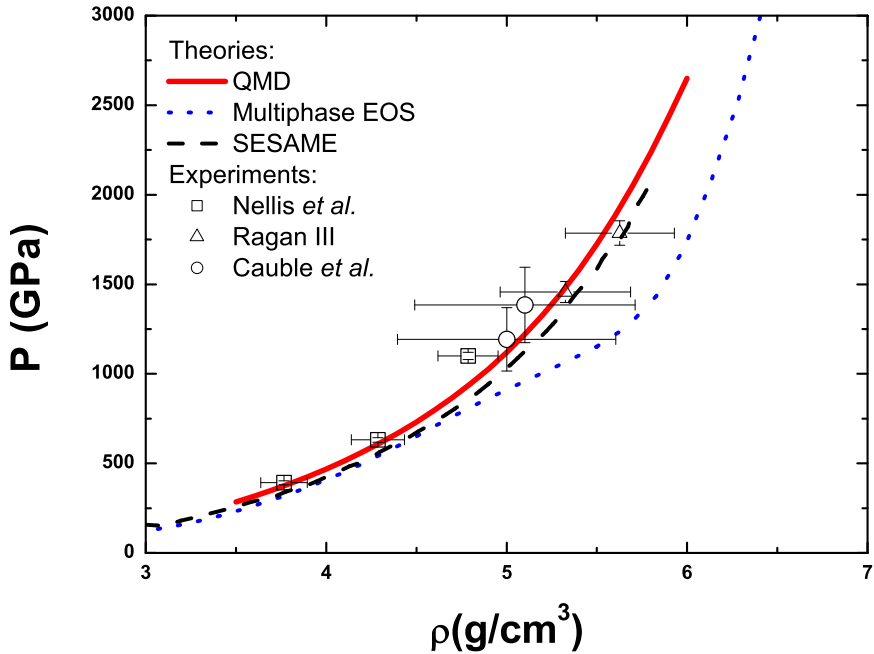}
\caption{(Color online) Hugoniot curve computed by QMD simulations
(red line) are compared with previous results. Underground nuclear
explosive experiments by Nellis \emph{et al.} \cite{Nellis1997} and
Ragan III \cite{Ragan1982} are labeled as open square and triangle.
High power laser results by Cauble \emph{et al.} \cite{Cauble1998}
are shown as open circles. SESAME \cite{Lyon1992} and multi-phase
EOS \cite{Benedict2009} are shown as dashed and dotted lines,
respectively.}\label{fig_hugoniot}
\end{figure}

The Hugoniot curve is shown in Fig. \ref{fig_hugoniot}, where
previous theoretical and experimental results are also provided for
comparison. In the warm dense region, the nature of the continuous
transition from condensed matter to dense plasma remains an
outstanding and interesting issue in high-pressure physics. One way
to address this issue is to measure EOS in the range of 0.1 to 5
TPa. The high shock pressures required to span this range have been
reached traditionally with strong shock waves generated by
underground nuclear explosives \cite{Ragan1982,Nellis1997}, then
accessed by high intensity lasers \cite{Cauble1998}. Good agreement
is found from Fig. Fig. \ref{fig_hugoniot} between our
QMD-determined EOS and those obtained experimentally. The present
QMD EOS indicates a smooth transition from condensed matter to
plasma at temperatures from 1.0 eV to 10.0 eV, where we do not find
any signs that suggest a first order plasma phase transition.
Theoretically, SEMASE EOS \cite{Lyon1992} also shows an overall
accordance with our results. On the contrast, the Hugoniot curve by
multi-phase EOS is soften at $\rho\sim6$ g/cm$^{3}$, which
corresponds to the imperfect join between the mixed theoretical
models \cite{Benedict2009}.

\subsection{Diffusion and viscosity}

\begin{figure}[!ht]
\includegraphics[width=12.0cm]{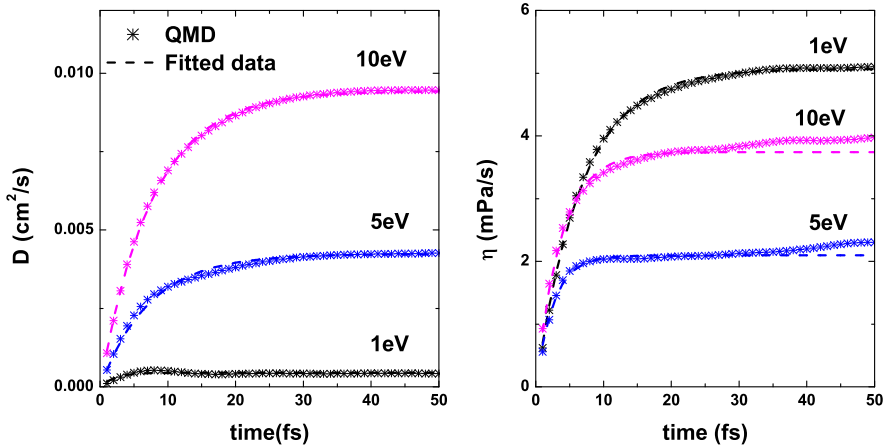}
\caption{(Color online) Self-diffusion coefficient and viscosity as
a function of time at a density of 5.0 g/cm$^{3}$ and different
temperatures of 1.0, 5.0, and 10 eV. The fits (dashed lines) are
performed for a sample window of
[0,4$\tau$-5$\tau$].}\label{fig_fit}
\end{figure}

We have performed \emph{ab initio} quantum-mechanical simulations
with the FT-DFT method to benchmark the dynamic properties of Be in
the WDM regime. An example of the QMD results for the self-diffusion
coefficient and viscosity of 5.0 g/cm$^{3}$ at temperatures of 1.0,
5.0, and 10.0 eV are displayed with their fits in Fig.
\ref{fig_fit}. The current simulations have the trajectory of
10$\mathtt{\sim}$20 ps and correlation times between 100 and 200 fs.
The computed error lies within 10\% for the viscosity. Due to the
fitting procedure and extrapolation to infinite time a total
uncertainty of $\mathtt{\sim}$20\% is estimated by experience. Since
the particle average gives an additional $\frac{1}{\sqrt{N}}$
advantage, the error in the self-diffusion coefficient is less than
1\%.

\begin{figure}[!ht]
\includegraphics[width=12.0cm]{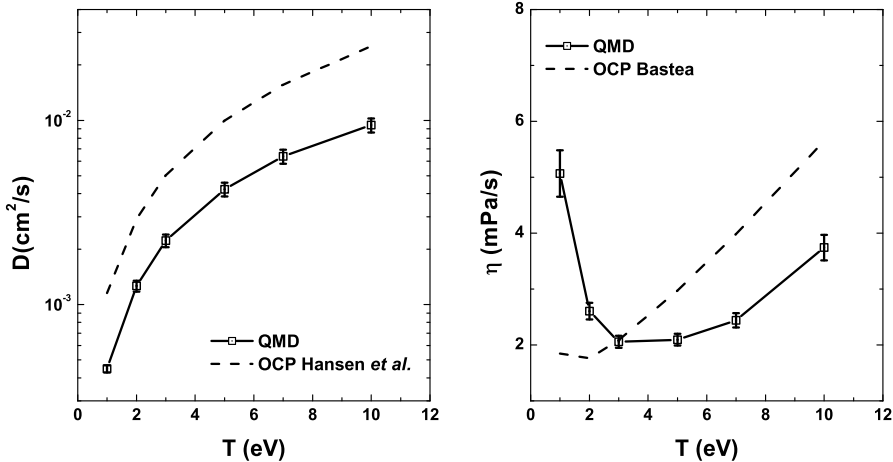}
\caption{(Color online) Self-diffusion coefficient (left panel) and
viscosity (right panel) as a function of temperature at a density of
5.0 g/cm$^{3}$. Only statistical error has been considered
here.}\label{fig_d_v}
\end{figure}

Idealized model, such as one-component plasma (OCP) model, concerns
the interaction through the Coulomb potential within a neutralizing
background of electrons. A large amount of molecular dynamics and
Monte Carlo simulations based on OCP model
\cite{Bastea2005,Lambert2007,Bernu1978,Donko1998,Daligault2006,Hansen1975}
have demonstrated that physical properties like diffusion and
viscosity can be represented in terms of coupling coefficient, which
is defined by the ratio of the potential to kinetic energy:
\begin{equation}\label{eq_gamma}
    \Gamma=\frac{Z^{2}e^{2}}{ak_{B}T},
\end{equation}
where $Ze$ is the ion charge, and $a=(3/4\pi n_{i})^{1/3}$ is the
ion-sphere radius with $n_{i}=\rho/M$ the number density. A
memory-function has been used by Hansen \emph{et al.}
\cite{Hansen1975} to analyze the velocity autocorrelation function
to obtain the diffusion coefficient for the classical OCP:
\begin{equation}\label{eq_dim_dif}
    \frac{D}{\omega_{p}a^{2}}=2.95\Gamma^{-1.34},
\end{equation}
with $\omega_{p}=(4\pi n_{i}/M)^{1/2}Ze$ being the ion plasma
frequency.

Bastea \cite{Bastea2005} has performed classical molecular-dynamics
simulations of the OCP and fits his results to the form
\begin{equation}\label{eq_dim_vis}
    \frac{\eta}{n_{i}M\omega_{p}a^{2}}=A\Gamma^{-2}+B\Gamma^{-s}+C\Gamma,
\end{equation}
with $s=0.878$, $A=0.482$, $B=0.629$, and $C=0.00188$. As the OCP
model is restricted to a fully ionized plasma, determination of the
ionization degree for WDM permits an extension of the OCP formulas
to cooler systems. A reasonable choice is to replace $Z$ in Eq.
(\ref{eq_gamma}) with an effective charge $\bar{Z}$. As a
consequence, we have introduced average-atom (AA) model
\cite{Rozsnyai1972}, which solves the Hartree-Fock-Slater equation
in a self-consistent field approximation assuming a finite
temperature. In the densities and temperatures we explore, the
effective charge $\bar{Z}$ is 2.0, which corresponds to the case of
full ionization of the $2s$ electrons.

QMD and OCP results for the self-diffusion coefficient, at the
density of 5.0 g/cm$^{3}$, are shown in the left panel in Fig.
\ref{fig_d_v}. The tendency for the self-diffusion coefficient with
respect to temperature is similar for QMD simulations and OCP model.
However, OCP model predicts a larger value of the diffusion
coefficient compared to QMD results ($\sim$1.5 times bigger).
Results for the viscosity, which consists contribution from
interatomic potential and kinetic motion of particles, are plotted
in the right panel of Fig. \ref{fig_d_v}. The minimum of viscosity
along temperature can be attribute to a combined effect, that is,
contribution from interatomic potential decreases with temperature,
while contribution from kinetic motion increases with temperature.
OCP model indicates the local minimum around 2.0 eV, and QMD
suggests the location around 3.0 eV.

\begin{figure}[!ht]
\includegraphics[width=12.0cm]{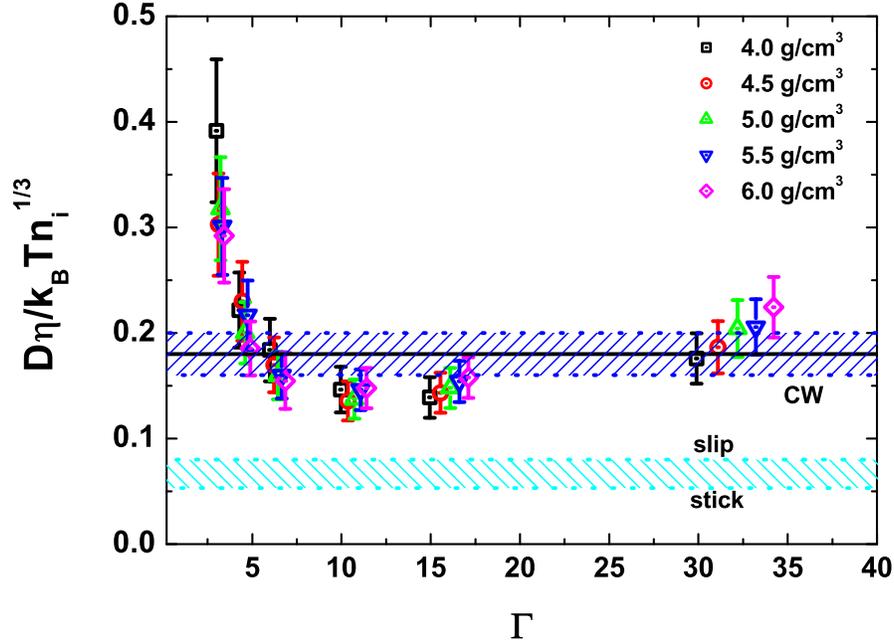}
\caption{(Color online) Examination of the Stokes-Einstein relation
along the coupling parameter in the warm dense region. Predictions
by Chisolm and Wallace \cite{Chisolm2006} are shown as blue region
and denoted as `CW' in the figure. The flat cyan lines show the
constant values of $C_{SE}$ for stick and slip boundary conditions
\cite{Einstein1908,Sutherland1905}.}\label{fig_SE}
\end{figure}

The Stokes-Einstein relation gives a connection between the
diffusion and shear viscosity:
\begin{equation}\label{eq_SE_relation}
    F_{SE}[D,\eta]=\frac{D\eta}{k_{B}Tn_{i}^{1/3}}=C_{SE},
\end{equation}
where $F_{SE}$ is a shorthand notation for the relationship between
the transport coefficients and $C_{SE}$ is a constant. Several
prescriptions \cite{Cappelezzo2007} for determining $C_{SE}$ are
available. Chisolm and Wallace \cite{Chisolm2006} have provided an
empirical value of 0.18 $\pm$ 0.02 from a theory of liquids near
melting. On the other hand, $C_{SE}$, which has been derived based
on the motion of a test particle through a solvent, was assumed to
range from $1/6\pi$ \cite{Einstein1908} to $1/4\pi$
\cite{Sutherland1905} depending on the limits of the slip
coefficient from infinity (stick) to zero (slip). Here, we have
examined the behavior of Be over the various regimes we explore. As
shown in Fig. \ref{fig_SE}, we have plotted the Stokes-Einstein
expression $F_{SE}(D,\eta)$ as a function of the coupling parameter
$\Gamma$ using the diffusion coefficients and viscosities from QMD
simulations at densities of 4 $\sim$ 6 g/cm$^{3}$ and temperatures
from 1 $\sim$ 10 eV. As the expected fitting error of $\sim$ 20\%
for the viscosity, the QMD results are bounded by the classical
values of $C_{SE}$ from below (slip limit) and the Chisolm-Wallace
liquid metal value from above at temperatures below 7.0 eV
(corresponding to $\Gamma>4.0$). The function $F_{SE}[D,\eta]$ at
the higher temperature evinces a sharp increase with temperature.
The near-linear rise of the diffusion coefficient with temperature
in the whole region basically cancels the temperature dependence of
the denominator. As a consequence, the behavior of $F_{SE}[D,\eta]$
is dominated by the viscosity, which rises sharply at high
temperatures as shown in Fig. \ref{fig_d_v}. As mentioned above,
this abrupt bend in the viscosity with temperature reflects a change
from potential to kinetics dominated regimes.

\subsection{Lorenz Number}

\begin{figure}[!ht]
\includegraphics[width=12.0cm]{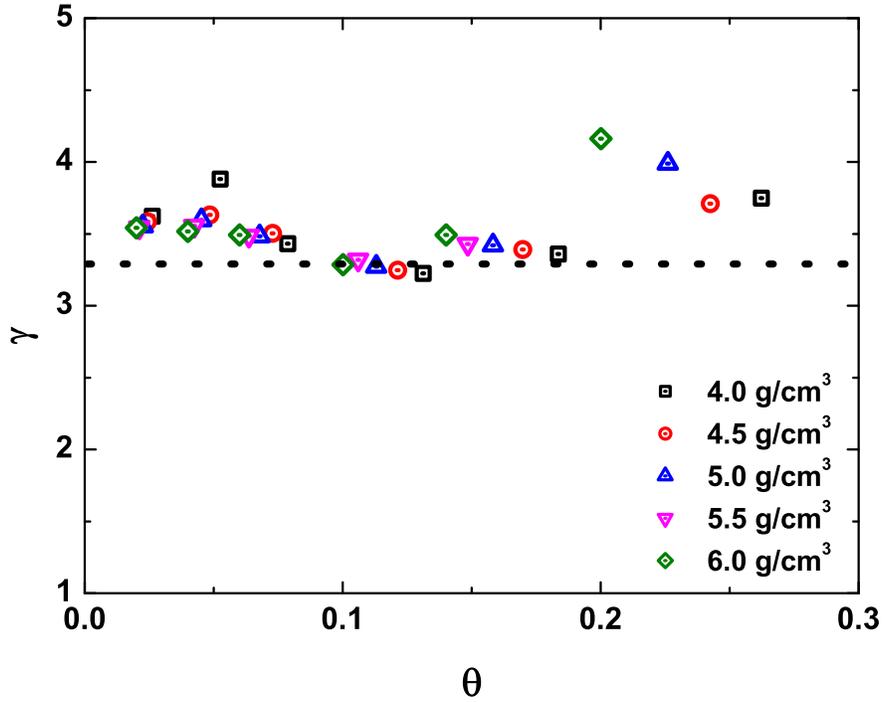}
\caption{(Color online) Calculated Lorenz number $\gamma$ as a
function of degenerate parameter $\theta$. The ideal Sommerfeld
number is plotted as dotted line in the figure.}\label{fig_lorenz}
\end{figure}

We have also computed the Lorenz number defined as
\begin{equation}\label{eq_lorenz}
    L=\frac{K}{\sigma T}=\gamma\frac{e^{2}}{k_{B}^{2}},
\end{equation}
where $K$ and $\sigma$ are the thermal and electrical
conductivities, respectively. $\gamma$ depends on the screened
potential and corresponds to the scattering of the electrons
\cite{Chester1961}. In a degenerate ($\theta\leq1$) and coupled
plasma ($\Gamma\geq1$), $L$ or $\gamma$ is a constant ($\pi^{2}/3$),
reaching the ideal Sommerfeld number, which is the value valid for
metals, and the Wiedemann-Franz law is recovered in an elastically
interacting electron system. As temperature is very high, the WDM
enters into a nondegenerate case ($\theta\gg1$ and $\Gamma\ll1$),
the Lorenz number reaches the value of kinetic matter (4 or 1.5966
depending on the $e$-$e$ collisions). In the intermediate region, no
assumptions can be found for predicting the Lorenz number and one
cannot deduce the thermal conductivity from the electrical
conductivity by using Wiedemann-Franz law.

In Fig. \ref{fig_lorenz} we show the behavior of $\gamma$ as a
function of the degeneracy parameter $\theta$
($\theta=k_{B}T/E_{F}$, with Fermi energy
$E_{F}=\hbar^{2}\frac{(3\pi^{2}n_{e})^{2/3}}{2m_{e}}$ and electron
density $n_{e}$). Here, we should stress that in QMD simulations,
the electrical conductivity $\sigma$ and electronic thermal
conductance $K$ can be directly evaluated without using any
assumption of Lorenz number, which is highly dependent on the two
non-dimensional parameters $\theta$ and $\Gamma$. In the present
warm dense regime, the Lorenz number vibrates around the Sommerfeld
limit at low temperatures, and as $\theta$ increases, a departure of
the Lorenz number from the ideal value can be observed from Fig.
\ref{fig_lorenz}.

\section{CONCLUSION}
\label{sec-conclusion}

In the present work, clear chains have been demonstrated in
investigating the thermo-physical properties of warm dense Be. The
EOS has been calculated through \emph{ab initio} molecular dynamic
simulations, and smooth functions have been constructed to fit the
QMD wide range EOS data, which show good agreement with the
underground nuclear explosive and high pulsed laser experimental
results. Based on Green-Kubo relation, the self-diffusion
coefficients and viscosity have then been determined, and as a
reference, OCP model with an effective charge determined from the
average-atom model has also been employed. A Stokes-Einstein
relation between the viscosity and diffusion coefficient holds the
general feature of liquids as predicted by Chisolm and Wallace in
the strong coupling region ($\Gamma>4$), while, a sharp increase has
been observed as temperature arises. The Kubo-Greenwood formula
provides an efficient way to study the electrical conductivity and
electronic heat conductance in the warm dense regime. Through QMD
simulations we have showed that the Wiedemann-Franz law is satisfied
for the degenerate regime. Our present results are expected to shed
light on the hydrodynamic modeling of target implosions in ICF
design.

\begin{acknowledgments}
This work was supported by NSFC under Grants No.11275032, No.
11005012, and No. 51071032, by the National Basic Security Research
Program of China, and by the National High-Tech ICF Committee of
China.
\end{acknowledgments}

\end{document}